\documentclass{aa}
\usepackage{graphicx}
\usepackage{txfonts}
\usepackage{natbib}
\begin{document}
\authorrunning{O.~Pani\'c et al.}
\titlerunning{Warm molecular gas and kinematics in the disc around HD~100546}

\title{Observations of warm molecular gas and kinematics \\ 
in the disc around HD~100546}
\author{O. Pani\'c\inst{1,}\inst{2} \and E.~F.~van Dishoeck\inst{1,}\inst{3} \and M.~R.~Hogerheijde\inst{1} \and A.~Belloche\inst{4} \and R.~G\"usten\inst{4} \and W.~Boland\inst{1,}\inst{5} \and A.~Baryshev\inst{6}}
\institute{Leiden Observatory, Leiden University, P.O.Box 9513, 2300 RA, Leiden, The Netherlands 
\and
European Southern Observatory, Karl Schwarzschild str. 2, 85748 Garching, Germany \\
\email{opanic@eso.org}
\and
Max-Planck-Institut f\"ur extraterrestrische Physik, Postfach 1312, 85741 Garching, Germany
\and 
Max-Planck Institut f\"ur Radioastronomie, Auf dem H\"ugel 69, D-53121 Bonn, Germany
\and
Nederlandse Onderzoeksschool Voor Astronomie (NOVA), P.O.Box 9513, 2300 RA, Leiden, The Netherlands
\and 
SRON Netherlands Institute for Space Research , PO Box 800, 9700 AV Groningen, The Netherlands
}
\date{Received  ; Accepted }
 
\abstract
{The Herbig Ae/Be star \object{HD~100546} hosts one of the most extensively studied discs in the southern sky. However, not much is known about this disc's gas and large scale kinematics.}
{Our aim is to establish whether the disc is gas-rich, and to study the disc temperature and kinematics.}
{We detect and study the molecular gas in the disc at spatial resolution from 7$\farcs$7 to 18.9$\farcs$ using the Atacama Pathfinder Experiment telescope\footnote{This publication is based on data acquired with the Atacama Pathfinder Experiment (APEX). APEX is a collaboration between the Max-Planck-Institut f\"ur Radioastronomie, the European Southern Observatory, and the Onsala Space Observatory.}. The lines $^{12}$CO $J=$7--6, $J=$6--5, $J=$3--2, $^{13}$CO $J=$3--2 and [C I] $^3$P$_2$--$^3$P$_1$ are observed, diagnostic of disc temperature, size, chemistry, and kinematics.
We use simple parametric disc models and a molecular excitation and radiative transfer code to fit the observed spectral line profiles.}
{Our observations are consistent with more than 10$^{-3}$~M$_{\odot}$ of molecular gas in a disc of $\approx$400~AU radius in Keplerian rotation around a 2.5~M$_{\odot}$ star, seen at an inclination of 50$\degr$. The detected $^{12}$CO lines arise from gas at 30-70~K.  Asymmetry in the $^{12}$CO line emission suggests that one side of the outer disc is colder by 10-20~K than the other, and is not due to pointing offsets, cloud absorption and asymmetry in the disc extent. }
{Both low- and high-$J$ $^{12}$CO lines are dominated by the outermost disc regions, and indicate a 400~AU radius. The $^{12}$CO $J=$6--5 line arises from a disc layer higher above disc midplane, and warmer by 15-20~K than the layer emitting the $J=$3--2 line. The existing models of discs around Herbig Ae stars, asuming a B9.5 type model stellar atmosphere overproduce the [CI] $^3$P$_2$--$^3$P$_1$ line intensity from HD~100546 by an order of magnitude.}

\maketitle

\section{Introduction}\label{s:intro}

Over the past decade our understanding of the structural and physical properties of discs around young stars has increased from basic theoretical modelling of the spectral energy distributions (SEDs) constrained by observations with no spatial information, to modelling based on not only the SEDs, but also spatially resolved dust observations, like scattered light images and interferometry \citep{pinte,panic2008,tannirkulam}. Two decades ago, the first submillimetre interferometer observations resolved the molecular gas emission spatially and this allowed major progress in understanding the disc kinematics, structure and chemistry \citep[e.g.,][]{beckwith1987,koerner1993,dutrey1994}. 
{A more recent example is a study of one of the brightest discs around a Herbig~Ae star, \object{HD~163296} shown in \citet{isella}.}
(Sub)millimetre gas and dust emission is the ideal probe of the global disc properties, like size, mass and radial distribution of disc material, because the bulk of the disc mass is located beyond 100~AU from the star, at temperatures of 10-50~K that dominate this part of the spectrum. 
Disc models which include constraints of both dust and molecular gas observations have stressed the importance of analysing the gas and dust components simultaneously, in the context of a {single disc model with a three-dimensional temperature and density structure} \citep[][]{wilner,raman,panic2008,panic2009}. 

Until recently, observations of rotational transitions of molecules in the submillimetre regime were focused primarily on the low-$J$ emission from $^{12}$CO, up to the $J=$3-2 line \citep{greaves2000,thi2001,qi2004,thi2004,dent2005}. In two of the brightest and most studied sources, \object{TW~Hya} and \object{LkCa~15}, the observations of higher-$J$ transitions of $^{12}$CO, up to J$=$6--5 ($E_{\rm k}=$116~K), were compared to the low-$J$ lines, providing estimates of the gas temperature in the intermediate-height molecular layer \citep{vzadelhoff}, crucial ingredients for chemical modelling of discs. These single-dish line spectra were fitted using simplistic disc models, deriving a temperature of 20-40~K in the $^{12}$CO line emitting layers of LkCa~15, and more than 40~K in TW~Hya. 
\citet{qi2006} analysed submillimetric interferometer observations of TW~Hya in the context of a disc structure based on an accretion disc model \citep{calvet2002}. Based on $^{12}$CO $J=$6--5, 3--2 and 2--1 observations, they show that X-ray heating of the gas is efficient in this source, in addition to the stellar radiation field. Such diagnostics of gas heating and ionisation improve our understanding of how the gas content evolves in discs.

The emerging (sub)millimetre facilities in the Southern hemisphere like the Atacama Pathfinder EXperiment (APEX) and the Australia Telescope Compact Array (ATCA) are opening a window towards the star-forming regions of the Southern sky and are well suited to study circumstellar disc emission. These instruments also pave the path for future observations with the Atacama Large Millimetre / Submillimeter Array (ALMA), which will drastically improve our knowledge of disc structure and evolution \citep[e.g.][]{GD2008}. We use APEX receivers APEX-2a and CHAMP$^+$ to observe the $^{12}$CO $J=$7--6, $J=$6--5, $J=$3--2, $^{13}$CO $J=$3--2 and [C I] $^3$P$_2$--$^3$P$_1$ line emission towards the disc around the young intermediate-mass star HD~100546. A wealth of observations of dust in this bright disc \citep{waelkens1996,malfait,grady2001,augereau,bouwman2003,acke2006,ardila2007} has motivated us to probe its molecular gas content and kinematics. The chosen transitions are particularly sensitive to the gas in the warm upper layers and kinematics of the outer disc. Our millimetre line observations probe the outer radius and inclination. The existing observational constraints on these parameters in the disc around HD~100546 provide an excellent basis for the analysis of our data. Our observations also provide a bridge toward even higher-$J$ far infrared $^{12}$CO transitions to be observed with the Herschel Space Observatory.
HD~100546 is a young B9V type, 2.5~M$_{\odot}$ star, classified as a Herbig Be star due to its isolation, infrared excess and silicate emission \citep{the1994,malfait}. With a distance of 103$\pm$6~pc, measured by Hipparcos, this is one of the nearest Herbig Ae/Be stars. The age of the star is estimated to be greater than 10~Myr \citep{vdancker}. This makes the presence of circumstellar material intriguing, considering that discs are found to dissipate within 10~Myr in most young stars \citep[e.g.,][]{hollenbach2000,hernandez2007,hillenbrand2008}. 
Based on SED modelling,
\citet{bouwman2003} postulate the presence of an inner hole in the
disk with a 10~AU radius, possibly caused by a Jupiter-sized planet
\citep[see also][]{acke2006}.
Direct evidence of cold disc material at larger radii is provided by ATCA observations of \citet{wilner} at 89~GHz (3.4~mm) and 2$\arcsec$ resolution, with the flux of 36$\pm$3~mJy, values consistent with the 1.3~mm observations of \citet{henning1998}. They do not detect HCO$^+$ $J=$1--0 line emission and speculate that photodissociation of its progenitor species $^{12}$CO, in the upper disc layers or an overall gas depletion may be the reason for this. Recent spectroastrometric observations of rovibrational $^{12}$CO transitions by \citet{vdplas} suggest that this molecular species is missing from the inner disc, at least up to 8~AU from the star \citep[also see][]{brittain2009}. 

Scattered light imaging of HD~100546 reveals the disc extending up to 4$\arcsec$ from the star viewed at an inclination of 50$\degr$, and an interesting disc structure resembling spiral arms \citep{pantin,augereau,grady2001}. This structure was interpreted as due to disc perturbation by a companion \citep{quillen2005} or a warped disc structure \citep{quillen2006}.  
Coronographic imaging by \citet{augereau} shows steep surface brightness profiles in the environment of HD~100546 indicative of optically thin emission in the near-infrared, with surface densities as low as 10$^{-3}$~g~cm$^{-2}$. Their images trace the emission of small dust ($<5\,\mu$m), extending out to 800~AU from the star. The authors suggest the presence of an optically thick disc with a 400~AU radius, and an optically thin flattened halo or envelope farther from the star.  
The scattered light images hint at the presence of gas in the disc that supports the disc vertical structure. 

In this work, we detect and study the molecular gas, its kinematics and temperature, in the disc at spatial resolution from 7$\farcs$7 to 18.9$\farcs$.
In Sect.~\ref{s:obs&res} we present our observations of $^{12}$CO, $^{13}$CO and [C I] lines. All $^{12}$CO lines are detected, there is a tentative detection of the $^{13}$CO line, while [C I] emission is not detected. We model the spectra in
Sect.~\ref{s:discussion} discussing the implications for the disc size, mass and kinematics. We identify the regions dominating the observed lines, and derive their temperatures. Section~\ref{s:conclusions} summarises our results.

\section{Observations and results}\label{s:obs&res}

The observations of $^{12}$CO $J=$6--5 at 691.472 GHz and [C I]
$^3$P$_2$--$^3$P$_1$ ([C I] $J=$2--1 hereafter) at 809.344 GHz towards HD~100546 at R.A.$=$11$^\mathrm{h}$33$^\mathrm{m}$25$\fs$4 and Dec$=-$70$\degr$11$\arcmin$41$\arcsec$ (J2000) were obtained simultaneously, using the
CHAMP$^+$ heterodyne array receiver \citep{gusten} on APEX on 2008 November 11. The 7 pixels in each wavelength band are
arranged in a hexagon of 6 pixels around one central pixel pointed towards the source, with beam sizes of 9$\arcsec$ at 691~GHz and 7$\farcs$7 at 809~GHz. {The data were obtained in the position switching mode in which the telescope moved between the on-source position and a reference position at an equatorial offset of (-1000$\arcsec$, 0$\arcsec$) from the source. }
The backend consisting of the Fast Fourier Transform
Spectrometer unit was used on all pixels, providing a spectral resolution of 0.37~MHz or 0.14-0.16~km~s$^{-1}$ at these frequencies, and covering a bandwidth of 1.5~GHz or 4096 channels.
Main beam efficiencies are 0.40 at 691~GHz and 0.37 at 809~GHz. Calibration is uncertain by $\approx$30$\%$ at both frequencies. Pointing was performed directly prior to, and after the on-source integration providing an accuracy better than 3$\arcsec$. The pointing source was 07454-7112, about 20$\degr$ away from HD~100546, at the same airmass. The CO $J=$6--5 line was also observed on 2008 November 10 in jiggle mode and its intensity and spectral profile were found to be
the same within 20\%. Since these data are noisier, the data sets taken on different dates were not combined. During this latter observation, the high band was tuned
to $^{12}$CO $J=$7--6 at 806.665 GHz, with the beam efficiency of 0.37. Only the high S/N $^{12}$CO $J=$6--5 data taken on 2008 November 11 are used in the further analysis.
 
The $^{12}$CO $J=$3--2 line at 345.796~GHz, and the $^{13}$CO $J=$3--2 line at 330.588~GHz, were observed on 2005 July 27 and 28 with the APEX-2a receiver using a single pointing. The channel spacing of these data is 61~kHz or 0.05-0.06~km~s$^{-1}$, and the spectral resolution 98~kHz or 0.09~km~s$^{-1}$ \citep{klein}. The main beam efficiency of APEX at 346~GHz is 0.73, and the beam sizes are 18$\farcs$1 and 18$\farcs$9 at 345.796~GHz and 330.588~GHz, respectively. {Calibration is uncertain by $\approx$20$\%$ at these frequencies.} Our $^{12}$CO $J=$3--2 line data were presented in \citet{panicHAes}. In all our observations, APEX forward efficiency of 0.95 was taken into account.

\begin{table}
\caption{\label{tab2} Integrated intensities of the observed spectral lines.}
\centering
\begin{tabular}{c c c}
 \hline\hline
 Spectral Line & $I$\tablefootmark{a} & $FWHM$ \\

 & (K~km~s$^{-1}$) &(km~s$^{-1}$) \\
\hline
$^{12}$CO $J=$7--6  & 12.9$\pm$1.9 & 4.2 \\
$^{12}$CO $J=$6--5  & 17.7$\pm$0.9 & 4.2 \\
$^{12}$CO $J=$3--2  & 4.0$\pm$0.6 & 4.0 \\
$^{13}$CO $J=$3--2& 1.3$\pm$0.6\tablefootmark{b} & (...)  \\
CI $J=$2--1 &  $<$1.0\tablefootmark{c} & (...)  \\
\hline
\end{tabular}
\tablefoot{
\\
\tablefoottext{a}{The line intensity $I = \int{T_{\rm MB}\,dV}$ is integrated over the velocity range 0--10~km~s$^{-1}$.}\\
\tablefoottext{b}{Marginal 2$\sigma$ detection.}\\
\tablefoottext{c}{Upper limit given by the 2$\sigma$ value.}
}
\end{table}

\begin{figure}
\centering
\includegraphics[width=9.cm]{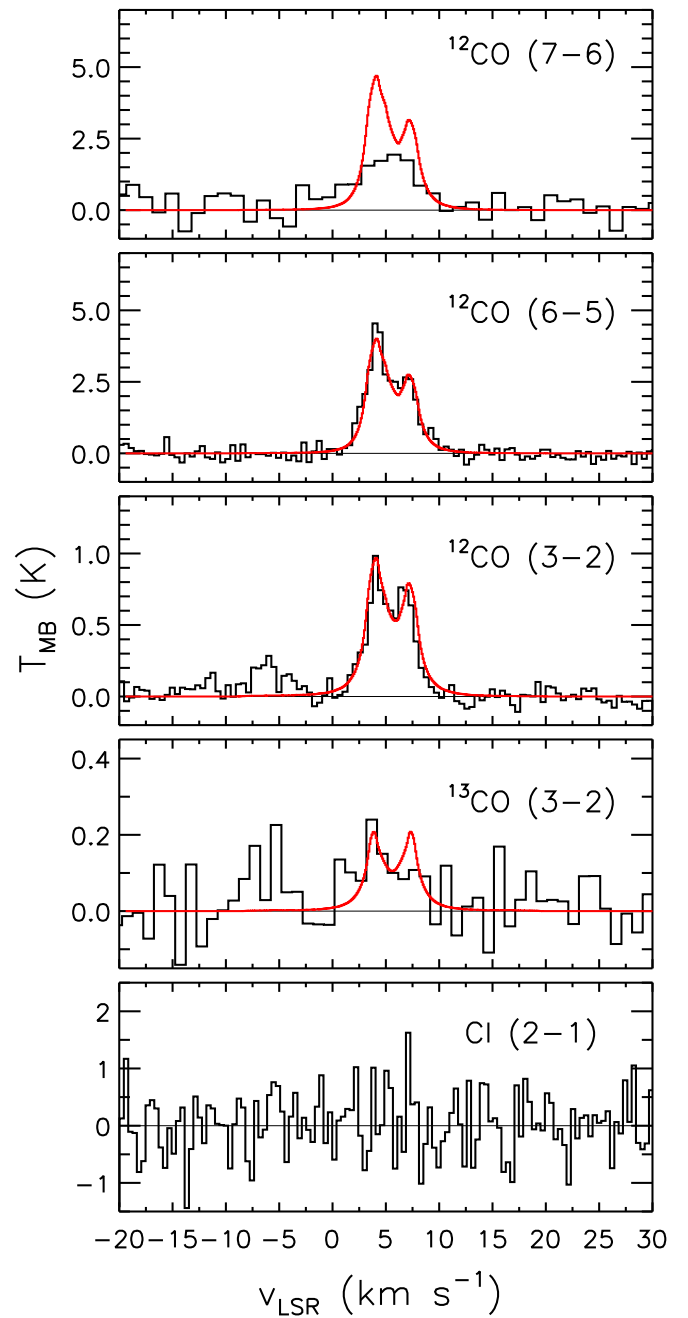}
\caption{\textit{From top to bottom:} Spectra of the $^{12}$CO $J=$7--6, $J=$6--5, $J=$3--2, $^{13}$CO $J=$3--2 and CI $J=$2--1 lines observed towards HD~100546 shown with black lines plotted as a histogram. The $^{12}$CO and $^{13}$CO spectra computed from our models of the line emitting layers of the disc (see Sections~\ref{ss:3265} and~\ref{ss:1376}) are shown with continuous red lines. At 100~AU from the star, our vertically isothermal models of the $^{12}$CO line emission have temperatures $T_{100}$=60-70~K, while for the $^{13}$CO a lower temperature $T_{100}$=30~K is used.}
\label{spectra}
\end{figure}

The data were reduced and analysed using the CLASS and STARLINK software. We detect all the observed $^{12}$CO line transitions at 7--20$\sigma$ in terms of the integrated intensity. The $^{13}$CO J$=$3--2 line is marginally detected at 2$\sigma$, and the [C I] $J=$2--1 line is not detected. {Figure~\ref{spectra} shows the observed spectra, corrected for the beam efficiency and re-binned to a lower spectral resolution. A first degree polynomial fit, except for the 0-10~km/s range in which the lines are emitted, is subtracted from the full spectral range for each data set.} The $^{12}$CO $J=$3--2 and $J=$6--5 lines are detected at the highest signal to noise ratio, and show a double peaked profile characteristic of disc rotation. The intensities integrated over the velocity range 0-10~km~s$^{-1}$, over which line emission is detected, are listed in Table~\ref{tab2} together with the full width at half-maximum (FWHM) of the lines with sufficiently well defined profiles.

In addition to the observations towards the source, the CHAMP$^+$ array provides measurements at nearby offsets. This setup provides an excellent way to discern the emission from the disc, with an estimated size of at least 400~AU in radius from the surrounding material known to be present further away from the star {\citep{henning1998,grady2001,augereau}}. The 400~AU lower limit to the size of the gas disc is set by the scattered light observations \citep{augereau} is close to the spatial resolution of our data. The central pixel of our CHAMP$^+$ data probes the $^{12}$CO $J=$6--5 line emission from the region of 9$\arcsec$ centered on the position of the star (460~AU radius), while the surrounding pixels probe the more distant regions (at a distance 1000-2000~AU from the star). Similarly, at the frequency of the $^{12}$CO $J=$7--6 line a smaller region around the star of 7$\farcs$7 is probed with the central pixel (400~AU radius), and regions roughly 1000-2000~AU from the star with the surrounding pixels. Table~\ref{tab3} provides an overview of the pixel positions and the corresponding fluxes integrated over a velocity range from 0 to 10~km~s$^{-1}$ velocity range, over which the $^{12}$CO lines are firmly detected in the central pixel. 
Compared to the on-source fluxes, these measurements show that the $^{12}$CO emission from the surrounding material is about 20 times weaker than from the region within 400~AU from the star. The $^{12}$CO $J=$3--2 line was observed with a single pointing, and the beam of 18$\farcs$1 is large enough to include any emission from regions beyond 400~AU. However, the strong resemblance in the line profile suggests that both low-$J$ and high-$J$ lines arise from the disc and that any contribution to the line emission by an extended low-temperature and low-velocity component is negligible.

\begin{table*}
\caption{\label{tab3} The $^{12}$CO $J=$6--5 and $J=$7--6 integrated line intensities at offset positions.}
\centering
\begin{tabular}{c c c | c c c}
 \hline\hline
 & $^{12}$CO $J=$6--5 & & & $^{12}$CO $J=$7--6 & \\
\hline 
RA Offset & Dec Offset & $I_{\mathrm{CO(6-5)}}$ & RA Offset & Dec Offset & $I_{\mathrm{CO(7-6)}}$ \\
($\arcsec$) & ($\arcsec$)  & (K~km~s$^{-1}$) & ($\arcsec$) & ($\arcsec$)  & (K~km~s$^{-1}$)\\
\hline 
-9.8  &  -17.0 & $0.4 \pm 0.2$ & +10.2 & -17.0 & $< 2.0$ \\
+8.8  &  -18.0 & $0.5 \pm 0.2$ & +20.2 & -1.4 &  $< 2.0$ \\
-18.9 &  -0.2  & $< 0.2$ & -9.0 & -16.5 &  $< 2.0$ \\
+19.4 &  -1.2  & $< 0.2$ & +11.0 & +16.2 &  $< 2.0$ \\
-9.3  &  +16.6 & $< 0.2$ & -18.8 & +0.2 &  $< 2.0$ \\
+10.0 &  +16.6 & $< 0.2$ & -9.1 & +16.9 &  $< 2.0$ \\
\hline
\end{tabular}
\tablefoot{
The line intensities $I = \int{T_{\rm MB}\,dV}$ are integrated over a velocity range 0-10~km~s$^{-1}$. The upper limits are given by the 2$\sigma$ value.
}
\end{table*}

Our data testify to the presence of warm molecular gas at distances less than roughly 400-500 AU from the star, corresponding to the smallest beam sizes in our observations (7$\farcs$7 and 9$\arcsec$). We obtain the following integrated intensity ratios, corrected for beam dilution (scaled to the same beam): {$^{12}$CO lines (6--5)/(3--2)=1.1$\pm$0.6 and ($^{12}$CO 3--2)/($^{13}$CO 3--2)$\geq$2.8$\pm$2.2. The errors include both the rms and flux calibration uncertainty.} 

There is a clear asymmetry in the profile of the $^{12}$CO $J=$3--2 and $J=$6--5 lines,
observed at a high signal-to-noise ratio, characterised by a 20-40\% difference in the line intensity between the two peaks. This asymmetry is certain, given that the spectra shown in Fig.~\ref{spectra} have been smoothed to a lower spectral resolution, and that each spectral bin contains an average of several measurements.

In some sources, asymmetries
like this are explained through confusion with foreground cloud emission, as seen
in $^{12}$CO lines towards IM~Lup \citep{vankempen} and DL~Tau
\citep{simon}.
However, in our off-pixels of the CHAMP$^+$ array we see no significant extended emission. While a slight pointing offset in the 6--5 line observations towards a symmetric source may result in a line asymmetry,{ the persistance of line asymmetry in our 18$\farcs$1 resolution observations of the 3--2 line indicates that the asymmetry is due to the source properties.} Furthermore, the two observations were taken years apart and a systematic offset in both observations is highly unlikely. We therefore exclude pointing offset as a cause of the observed line asymmetry.
If one side of the disc extends slightly further out than the other (e.g., on the SE side), the increase in disc surface at that side will contribute to the line flux at one side of the spectral line causing line asymmetry. This may be a plausible explanation for the 3-2 line, but not the observed 6-5 line with a 9$\arcsec$ beam size, unless the asymmetry is within 450~AU. Such an asymmetry may affect the scattered light images of HD~100546, with the disc extending further to the SE than to the NW of the star. However this is not seen in the observations of \citet{augereau}. Furthermore, the additional emission from this region would dominate the frequencies closer to the line centre, and not the blue-shifted peak.

Asymmetry is seen in $^{12}$CO lines from
sources with a pronounced structural disc asymmetry, like HD~141569
\citep{dent}. Another possibility is a temperature asymmetry in the disc. In the following section we analyse and model the observed lines and investigate these scenarios.

\section{Discussion}\label{s:discussion}

\subsection{Line ratios as temperature and density diagnostics}\label{ss:emission}

\subsubsection{$^{12}$CO 3--2/$^{13}$CO 3--2 line ratio and disc density}

The submillimetre $^{12}$CO line emission from circumstellar
discs is generally optically thick and arises from warm upper disc
layers. These lines are particularly sensitive to the temperature of
these layers, and therefore to the stellar and external illumination
of the disc surface. Comparison to isotopologue line emission is often used to assess the optical thickness.
\textit{If the $^{13}$CO 3--2 emission is treated as a detection,} {we obtain the line ratio of ($^{12}$CO 3--2)/($^{13}$CO 3--2)=2.8$\pm$2.2.} 
We make a rough estimate of the column density needed to reproduce the derived ratio using the RADEX online
tool \footnote{{\tt http://www.strw.leidenuniv.nl/$\sim$moldata/radex.html}} \citep{vdtak}. For kinetic temperatures 20-40~K, H$_2$ densities 10$^6$-10$^8$~cm$^{-3}$, and an isotopic ratio [$^{12}$C]/[$^{13}$C]=77 {\citep{wilson}, the ($^{12}$CO 3--2)/($^{13}$CO 3--2) ratio derived above indicates the $^{12}$CO column density of roughly 10$^{18}$~cm$^{-2}$, and excludes lower column densities. Using a standard $^{12}$CO abundance of 10$^{-4}$ with respect to H$_2$, this translates to the gas column density larger than 10$^{22}$~cm$^{-2}$.} 

\subsubsection{$^{12}$CO 6--5/$^{12}$CO 3--2 line ratio and disc temperature}

We compare the line ratio {($^{12}$CO 6--5)/($^{12}$CO 3--2)=1.1$\pm$0.6} to the calculations done using RADEX, and assume that these two lines are dominated by the same region in the disc. This assumption is generally true up to 200-300~AU from the star \citep[see calculations done in][]{vanzadelhoff}. For H$_2$ densities of 10$^6$-10$^8$~cm$^{-3}$, and $^{12}$CO column densities of 10$^{17}$-10$^{19}$~cm$^{-2}$, the
($^{12}$CO 6--5)/($^{12}$CO 3--2) line ratio $\approx$1.1 is reproduced at temperatures of 60-100~K. If the disc is large and sufficiently cold, the assumption that the two lines arise from the same region is not necessarily true, as the 3--2 line may be dominated by colder material at radii larger than the 6--5 line. It is therefore uncertain whether the line ratio reflects the difference in the temperature between the two \textit{vertical} layers or in the temperature of the \textit{radial} regions emitting the lines. We explore this further in Sect.~\ref{ss:3265} using more sophisticated radiative transfer tools and parametric models of disc structure.

The observed CO 6--5/3--2
integrated intensity ratio of{ 1.1$\pm$0.6} (observations scaled by the beam size) is higher than the ratios
close to 0.5 found for the discs around the T~Tauri stars LkCa~15 and
TW~Hya by \citet{vzadelhoff} and \citet{qi2006}.  The
TW~Hya 6--5/3--2 ratio has been interpreted as proof that the gas
temperature is higher than that of the dust in the surface layers
where gas and dust are not thermally coupled. Both UV radiation and
X-rays have been invoked to provide the additional gas heating \citep{jonkheid2004,kamp,glassgold,nomura,gorti}, although for Herbig~Ae stars the X-rays can be neglected \citep{kamp2008}.

The higher ratio found for HD~100546 implies higher gas temperatures
than for the aforementioned disks around T Tauri stars. This is expected based on models where
most of the gas heating comes from UV radiation from the central star,
since the cooler stars have less UV radiation \citep[e.g.,][]{woitke}.  The gas temperature is also strongly affected by the PAH
abundance in the disc. PAH features are seen prominently in the HD
100546 mid-infrared spectrum \citep[e.g.,][]{malfait}, and have
been shown to be spatially extended across the disk \citep{geers}. In contrast, neither of the two T~Tauri disks for which the $^{12}$CO $J=$6--5 line has been detected show PAH
emission, implying typical PAH abundances a factor of 10--100 lower \citep{geers2006}.

For the specific case of Herbig Ae disks, \citet{jonkheid2007} have
computed the gas heating and chemistry as well as the resulting CO
6--5 and 3--2 line emission starting from a set of dust disk models
developed by \citet{DD2005}. Models with decreasing disk
mass from 10$^{-1}$ to 10$^{-4}$ M$_\odot$ and decreasing dust/gas
ratios from $10^{-2}$ to $10^{-6}$ (simulating grain growth and
settling) were investigated. The UV radiation field was taken to be that of a B9.5 star \citep[see Fig. 3 of][]{jonkheid2006},
very close to that expected for HD~100546. An accurate treatment of
the shape of the UV field at short wavelengths, $<$1100~\AA, is very
important for a correct calculation of the CO photodissociation and
atomic carbon photoionization rates, since the UV intensity of a B9
star is orders of magnitude weaker in this wavelength range than the
(scaled) standard interstellar radiation field. The PAH abundance is
taken to follow the dust/gas ratio, with an abundance of $10^{-7}$ for
a normal dust/gas ratio of 0.01.

The $^{12}$CO $J=$6--5/3--2 integrated intensity ratios
computed by \citet{jonkheid2007} summed over the full extent of the disk model
are remarkably close to unity for the entire range of disk parameters
investigated, consistent with our observations. The absolute $^{12}$CO and $^{13}$CO intensities for
model B2 (a disk with a mass of 0.01 M$_{\odot}$ with a standard
gas/dust ratio of 100, appropriate for HD~100546) are also within
40\% of the observed values when scaled to the same source
distance and beam size, indicating a good agreement between models and
observations.

\subsection{Parametric disc models and origin of $^{12}$CO line emission}\label{ss:model}

Submillimetre $^{12}$CO line emission is analysed using two different
modelling approaches in the literature. Disc physical models, like the
irradiated accretion disc models of \citet{dalessio,DD2005}, are especially
well suited when the emission is spatially resolved. In this way, the
three-dimensional structure of the disc can be investigated, for
example when transitions of different optical depths are observed
\citep{panic2008}.  For spatially unresolved observations, like those
presented here, simplistic models with a limited number of free
parameters are more appropriate to derive some basic constraints on
disc properties based on the line spectrum \citep[method employed in, e.g.,][]{dutrey1994,GD98,vzadelhoff,dartois,dent2005}. Here we go a step further than in Sect.~\ref{ss:emission} in analysing the observations in the context of such models.

\begin{figure*}
   \includegraphics[width=16.cm]{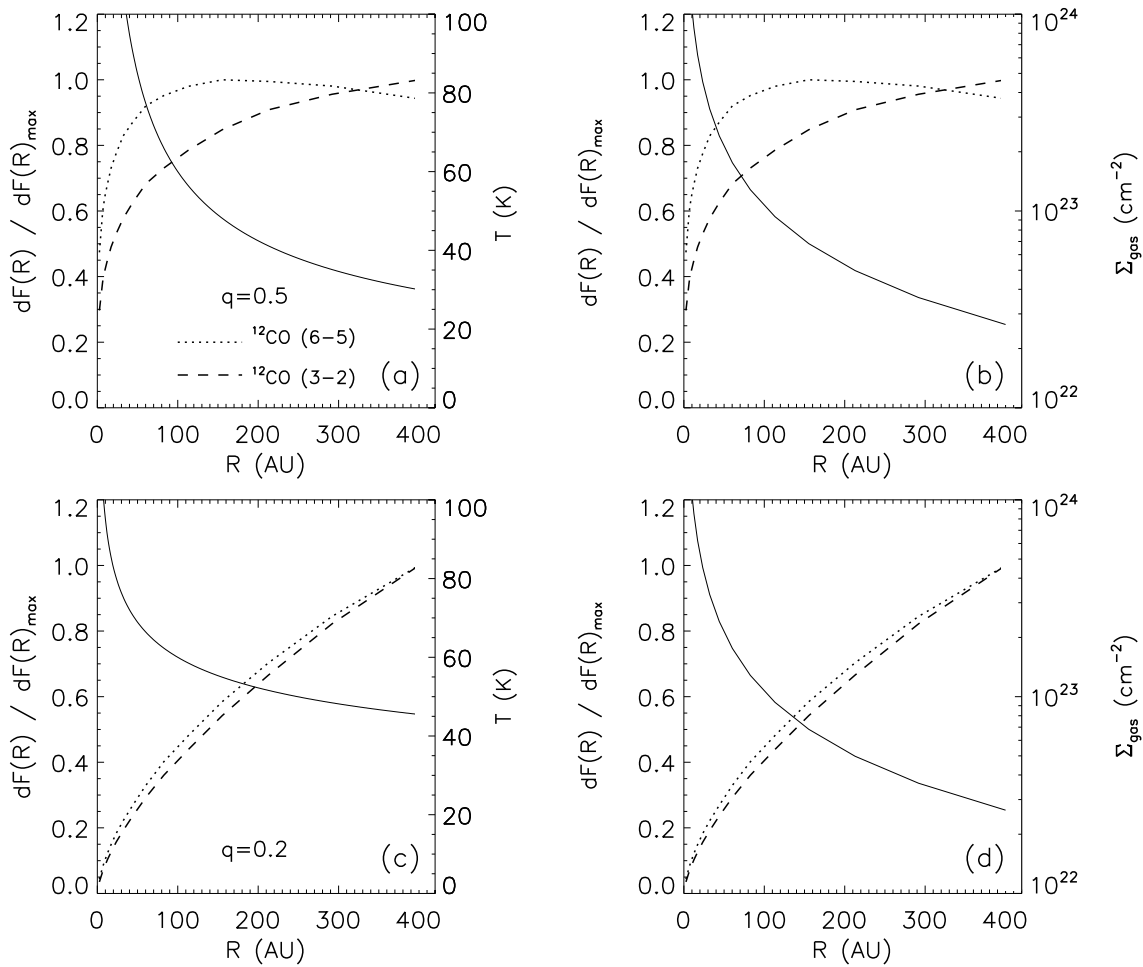}
\caption{{\textit{(a):} Radial temperature structure of a disc model from \citet{panicHAes}, with $R_{\rm out}=$400~AU, $T_{100}=$60~K and $q=$0.5 is shown with the black full line. The radial dependence of the line flux $dF(R)/dF(R)_{\rm max}$ for the $^{12}$CO 3--2 line and 6--5 line arising from this model is overplotted using the dashed and the dotted lines, respectively.\textit{(b):} The surface density of the same model is shown with the black full line, and line fluxes as in (a). 
\textit{(c) and (d):} Same as in (a) and (b), for the model with a shallower temperature profile, $q=$0.2 and its corresponding line fluxes as a function of radius.}}
\label{struct}
\end{figure*}

\textit{Density and temperature structure.} \citet{panicHAes} show that simple
power-law disc models, with a disc mass $M=0.01$~M$_{\odot}$, an H$_2$ surface density $\Sigma \propto R^{-1}$ and temperature $T=T_{100}~{\rm K}\,(R/{\rm 100\,AU})^{-q}$ with $T_{100}=$60~K and $q=$0.5, are a useful tool to
analyse low-$J$ $^{12}$CO transitions from discs around gas-rich Herbig~Ae stars. These models are vertically isothermal and have a vertical density structure in hydrostatic equilibrium. The temperature and surface density of such model with an outer radius of 400~AU is shown in Fig.~\ref{struct}, upper panels. The $^{12}$CO abundance with respect to H$_2$ is assumed to be 10$^{-4}$, constant throughout the disc.
Their analysis uses the assumption that the low-$J$ $^{12}$CO lines arise from the upper disc layers (optically thick) and that their fluxes are dominated by the outermost disc regions. Therefore the spectral profiles of these lines are very sensitive to the extent of the emission, with strong and narrow lines indicating a large disc and weak and relatively wider double-peaked lines characteristic of small discs. 

{
\textit{Spatial origin of the emission.} The regions of the disc probed by the $^{12}$CO $J=$6--5 and $J=$3--2 lines are studied in \citet{vanzadelhoff}. Their Fig.~6 shows the locations of the surfaces of opacity $\tau$=1 for these lines in a cross-section of their disc model. At distances up to 150~AU from the star, the two transitions trace similar depths into the discs, but at larger radii they trace different vertical layers: the low-excitation 3--2 line probes deeper and colder layers than the 6--5 line. To assess whether both 6--5 and 3--2 line fluxes are dominated by the same radial disc region, and to identify this region, we use the calculations of synthetic 6--5 and 3--2 spectra from disc models as described in the previous paragraph. We use two different temperature profiles, with $q=$0.5 and $q=$0.2. The radial temperature and surface density structure of the two models is shown in Fig.~\ref{struct}. In this exercise we use the face-on disc orientation, for simplicity, and calculate the emerging $^{12}$CO $J=$6--5 and $J=$3--2 line intensity along lines of sight at a number of different radial locations from the star. This is done using the molecular excitation and radiative transfer code RATRAN
\citep{hogerheijde} and CO-H$_2$ collision rates from the
Leiden Atomic and Molecular Database\footnote{http://www.strw.leidenuniv.nl/$\sim$moldata/} (LAMBDA;
\citealt{schoier}). 
We are primarily interested in knowing if there is a single radial region dominating the line flux more than other regions or the flux is distributed more evenly with the radius. Also, we investigate how the regions of strongest line flux spatially compare for the two $^{12}$CO transitions.
 
The contribution to the line flux from an annulus of a fixed width $dR$ located at a distance $R$ from the star is given by $dF(R)=2\,\pi\,I(R)\,R\,dR$, and therefore $dF(R)\propto I(R)\,R$.  The relative contribution of the annuli to the total line flux arising from the model $dF(R)/F_{total}$ is also proportional to $I(R)\,R$, allowing us to directly compare the importance of one radial region with respect to another.
The Figure~\ref{struct} shows $dF(R)$ for the $^{12}$CO $J=$6--5 (dashed line) and $J=$3--2 (dotted line) in case $q=$0.5 (upper panels) and $q=$0.2 (lower panels). For ease of comparison between the two lines, the $^{12}$CO $J=$3--2 and $J=$6--5 line fluxes are plotted with respect to their maximum value $dF_{\rm max}$. 
For both temperature profiles, the outermost disc region (200-400~AU) is very important, especially for the 3-2 line. For this line the strongest flux arises at the largest radii due to the efficient excitation at low temperatures found in the outer disc combined with the large surface area. This result is in line with our previous results in \citet{panicHAes} where the $^{12}$CO $J$=3--2 spectra are used as probes of disc sizes. While the 6-5 line is also sensitive to the outer disc regions, the upper panels of Fig.~\ref{struct} show that our model assumptions at 50~AU matter just as much as those at 400~AU. This is explained by the excitation energy of the $^{12}$CO $J$=6--5 line, that is somewhat high for the 30-40~K temperatures found in the outer disc in the model with a steeper temperature profile $q$=0.5. The flatter temperature distribution with $T >$40~K in the outer disc allows to efficiently populate the higher energy levels of $^{12}$CO, and in this case the outermost disc regions are clearly more important than 50-100~AU, due to the mere increase of the surface area with radius.
Understanding the diagnostic potential of the high-$J$ $^{12}$CO lines in discs is useful already for the analysis of spatially unresolved APEX observations (see Sect.~\ref{ss:3265}), but also shows that the future sub-arcsecond observations of high-$J$ $^{12}$CO lines in discs with ALMA could be used to probe the radial temperature structure in discs.
}

\textit{Disc mass.} Our initial choice of parameters, described above, roughly matches the observed $^{12}$CO $J=$6--5 and $J=$3--2 line fluxes, confirming the outer {radius} of 400~AU seen in scattered light observations. We performed test calculations for a lower value of the disc mass of 10$^{-3}$~M$_{\odot}$, i.e., ten times lower densities. The modelled line fluxes are found not to vary significantly with respect to the noise levels of our observations. Thus, in the disc mass range of 10$^{-3}$-10$^{-2}$~M$_{\odot}$, values found typically towards circumstellar discs, and temperatures similar to those assumed here, the observed  $^{12}$CO $J=$6--5 and $J=$3--2 lines are insensitive to disc mass. If we decrease the disc mass to allow the $^{12}$CO line emission in our models to be sensitive to the density variations in the midplane (i.e., emission dominated by the midplane), the disc midplane temperatures ($\leq$20~K in the outer disc) would be the only physically appropriate assumption for the temperature in the `low-mass' model. While the asymmetry in the line profile could be reproduced in this way, the resulting lines would be much weaker than observed. Given these considerations, we assume the two lines to be optically thick in the disc around HD~100546 and fix the disc mass to the initial value of 10$^{-2}$~M$_{\odot}$. {Our model is consistent with the thermal continuum emission of dust observed toward HD~100546 by \citet{henning1998,wilner}}. In the further analysis we fit the exact line shapes and intensities, with a particular focus on the line asymmetry seen in both $J=$6--5 and $J=$3--2 spectra.

\subsubsection{$^{12}$CO $J=$6--5 and $J=$3--2 line fit}\label{ss:3265}

The characteristic temperatures and masses derived in Sect.~\ref{ss:emission} for the disc around HD~100546 are similar to the models used in \citet{panicHAes}. The modelling of the $^{12}$CO 3--2 line of HD~100546 in \citet{panicHAes} yielded very rough estimates of the outer radius of 300~AU and inclination of 35$\degr$, consistent with the 400~AU and 50$\degr$ from the scattered light. We use their models to fit the $^{12}$CO spectra, with $T_{100}$ as free parameter and varying $q$ from its initial value of 0.5 where necessary. We fix the outer radius and inclination to the observationally constrained values 400~AU and 50$\degr$ \citep[see][and the discussion of disc size in Sect.~\ref{s:obs&res}]{augereau}. We use the Keplerian velocity field around a 2.5~M$_{\odot}$ star. The dependence of the spectral profile on the stellar mass in the range 2.0-3.0~M$_{\odot}$ does not affect our fit significantly.

The inner radius is assumed to be 0.6~AU, close to the dust sublimation radius. Although an inner hole of 13~AU is found in HD~100546, its presence would not affect our results because the molecular lines observed are dominated by disc regions far beyond the inner tens of AU. Because the observed $^{12}$CO lines trace warm molecular material ($>$20~K) and are insensitive to the colder regions deeper in the disc, it is reasonable to neglect the $^{12}$CO freeze-out.

As above, the line emission is calculated using RATRAN. The synthetic images from our disc models are convolved with the beam sizes of the corresponding observations, and the spectra are extracted towards the centre of the image. Where temperature asymmetry is modelled (see \ref{ss:3265}), pairs of synthetic images with different T$_{100}$ parameter were first combined - contributing each to a given side of the disc with respect to the semi-major axis. Following this, the combined image is convolved and spectra extracted.
Dust continuum emission,
although negligible for the molecular line transfer, is included in
the calculation and subtracted from the synthetic image cubes.

In the view of the result that the line emission appears to be insensitive to the density, the observed line asymmetry can not be caused by a density asymmetry in the disc, but it rather reflects an asymmetry in the disc temperature structure. To model such an asymmetry, we use the fact that the emission at the two peaks in the line profile is dominated by the two sides of the disc with respect to the minor axis.  
We assume that one side of the disc is colder than the other. This may be due to a combined effect of an asymmetry in disc illumination and geometry. We use
different $T_{100}$ parameters for the two sides of the disc and we explore different radial temperature distributions with $T \propto R^{-q}$, $q$=0-0.5. 

The best fit to the $^{12}$CO $J=$3--2 spectrum, seen in Fig.~\ref{spectra}, is obtained by assuming the radial temperature structure with $q$=0.5, as in \citet{panicHAes}, and $T_{\rm 100}$ of 60 and 70~K for the two disc sides, respectively. The line shape is consistent with the assumed Keplerian velocity field and disc inclination. The temperatures of 60-70~K compare well to the theoretical predictions of temperature in the regions where these lines saturate in discs \citep[see Fig.~6 in][]{vzadelhoff}. They show that at 100~AU from the star the $J=$6--5 and $J=$3--2 lines trace the same disc layer, we assume that the $^{12}$CO $J=$6--5 line is emitted at $T_{\rm 100}$ of 60 and 70~K, as found for the 3--2 line. We find the best fit to the $^{12}$CO $J=$6--5 spectrum, shown in Fig.~\ref{spectra}, assuming $T=$60~K$\times (R/100\,AU)^{-0.3}$ on one side of the disc and $T=$70~K$\times (R/100\;AU)^{-0.2}$ on the other. Thus, the radial temperature distribution in the 6--5 line emitting layer is flatter than in the 3--2 layer, consistent with the result of \citet{vanzadelhoff} for the outer disc regions where the 6--5 line traces warmer material higher in the disc surface than the 3--2 line. 

Therefore our $^{12}$CO line observations, although spatially unresolved, provide some information on the temperature radial distribution in the 6--5 and 3--2 line emitting layers and disc temperature asymmetry. It is important to note that slightly different sets of temperature values and slopes than those found here may provide an equally good fit to these observations. However, any such model requires higher outer disc temperatures to fit the 6--5 line than required for the 3--2 line, clearly showing that the two lines arise from different vertical layers in the outermost disc regions.

\begin{figure}
   \includegraphics[width=8.cm]{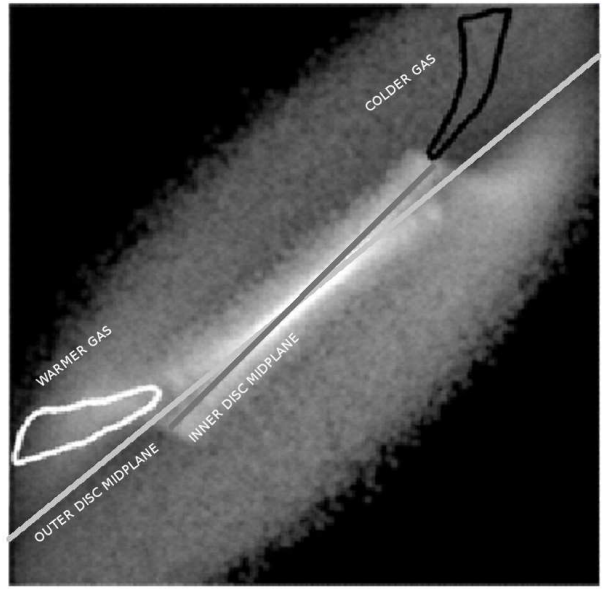}
\caption{Illustration of a warped disc structure, in an edge-on projection. Straight light and dark grey lines show the plane of the disc and of the inner warped disc component, respectively. The white curve shows the surface region with a higher temperature, heated well by the star, and the black curve shows the cooler region partly obscured by the warp. The image of the disc with a 400~AU radius is adapted from \citet{quillen2006}, where a detailed warped disc model is presented.}
\label{warp}
\end{figure}

The difference in temperature between the two sides of the disc may be explained by a warped inner disc, as illustrated in Fig.~\ref{warp}, with the elevated side of the inner disc intercepting a fraction of stellar light that would otherwise reach the outer disc, while the opposite side is slightly more illuminated, with the inner part of the disc shifted downwards. The possibility of an inner warp is suggested in \citet{quillen2006} for HD~100546, with an inner component extending up to 200~AU inclined by $\approx$15$\degr$ with respect to the outer component extending beyond that radius.

The temperature asymmetry is also possible if the disc has a different thickness at the two sides. This may happen in a disc with dust settling underway, if a planet or another body embedded in the disc stirrs the dust back up. In this case the `stirred' part of the disc intercepts more stellar light and becomes somewhat warmer. A companion body in HD~100546 is suggested in the literature, to explain the observed inner hole, gas kinematics, and spiral arms \citep{quillen2005,acke2006b}.

\subsubsection{$^{13}$CO $J=$3--2 and $^{12}$CO $J=$7--6 line fit}\label{ss:1376}

With the low signal to noise in our $^{13}$CO $J=$3--2 spectrum (marginal detection at 2$\sigma$), it is not possible to assess the {spectral} line shape. 
{The disc temperatures used to fit the $^{12}$CO lines in Sect.~\ref{ss:3265} overestimate the $^{13}$CO $J=$3--2 line intensity. A lower temperature of the $^{13}$CO $J=$3--2 emitting layers with respect to the $^{12}$CO $J=$3--2 (as discussed in Sect.~\ref{ss:1376}), a higher [$^{12}$C]/[$^{13}$C] isotopic ratio than the standard interstellar medium value of 77, or an overall decrease of CO abundance can explain our marginal detection
of the $^{13}$CO line. Freeze-out and/or selective
photodissociation processes in discs are known to affect the CO abundance and isotopic ratios. } Detailed modelling and spatially resolved
submillimetre line observations of $^{12}$CO and isotopologues would
allow to constrain the disc structure, evaluate the effect
of these processes \citep[,e.g.,][]{panic2008} and constrain the amount of molecular gas better.
{Clearly, there are a number of uncertainties involved in the interpretation of our $^{13}$CO $J=$3--2 data.  
For these reasons we limit the analysis to a comparison of our marginal detection to one simple model. A lower temperature of the $^{13}$CO emitting layers is expected due to a different optical depth of $^{13}$CO lines with respect to the $^{12}$CO.} In Fig.~\ref{spectra} we compare our data to model calculations done assuming low temperatures, $T_{100}$=30~K and $q$=0.5, an isotopic ratio [$^{12}$C]/[$^{13}$C]=77, and all remaining model parameters as in the simple axially symmetric power-law model shown in Fig.~\ref{struct} and described in Sect.~\ref{ss:3265}. Considering that the $^{13}$CO line emission probes colder and deeper disc layers, it is likely that temperature asymmetry derived for the upper layers from $^{12}$CO lines is negligible for $^{13}$CO. High-sensitivity observations of $^{13}$CO or C$^{18}$O submillimetre lines in the future may reveal symmetric line profiles.  

The $^{12}$CO $J=$7--6 and $J=$6--5 transitions arise from energy levels at $E_{\rm k}$ of 155~K and 116~K, respectively, and are expected to trace roughly the same disc layers. Figure~\ref{spectra} shows how our best-fit model of the 6--5 line compares to the observed 7--6 line. The observed line flux is about 50\% lower than the model prediction. This may be due to a slight mispointing of the telescope, as the size of the source (800~AU in diameter) is close to the 7$\farcs$7 beam size at this frequency. The noise levels of these observations are notably higher than in the 6--5 line observations and spectral profile not well defined. Due to these reasons, we do not fit the 7--6 line further but use the more reliable 6--5 line fit results in our analysis.

\subsection{Implications of the [C I] $J=$2--1 non-detection}

Figure 1 includes the high-quality spectrum around the [C I] $J=$2--1
line at 809.344~GHz. No significant feature is detected down to 0.3 K
rms in a 0.27~km~s$^{-1}$ velocity bin, implying a 2~$\sigma$ limit on the
integrated intensity of $\approx 1.0$~K~km~s$^{-1}$ over the range 0-10~km~s$^{-1}$ (same width as for the detected $^{12}$CO lines). Model B2
of \citet{jonkheid2007} predicts integrated [C I] intensities, scaled
to the distance of HD~100546, around 15--20~K~km~s$^{-1}$, whereas the
model intensities are even larger in disks with significant grain
growth and settling (BL model series in the abovementioned paper). Thus, while the CO data appear
entirely consistent with their sophisticated UV-heated disk atmosphere
models, the [C I] data are clearly discrepant by an order of
magnitude.

One possible solution could be that the radiation field contains more
carbon ionising photons than assumed here, shifting the chemical
balance from neutral to ionised atomic carbon. 
The predicted
[C II] line intensities for the model disks of \citet{jonkheid2007}
are very low, $<$0.1~K~km~s$^{-1}$, whereas they are more than an
order of magnitude higher for disks around T~Tauri stars with excess
UV emission \citep{jonkheid2004},{ consistent with an efficient carbon ionisation around Herbig Ae stars.} Such excess of UV emission could
come from the disk-star accretion boundary layer. Indeed, HD 100546 is
observed to undergo significant accretion, in spite of the known
(dust) gap in the inner disk \citep{vieira1999}. Also, the FUSE UV spectrum of HD 100546 shows significant flux, both
in the continuum and in lines such as O VI, in the wavelength range
where atomic carbon can be ionised \citep[][and G. Herczeg, private communication]{lecavelier2003}. The continuum emission should also
photodissociate CO, but not the stellar UV lines unless they overlap with
the discrete CO photodissociating transitions. Also, because CO is
self-shielding and the lines are optically thick, the CO emission may
be less affected than that of atomic carbon by any extra UV. However, detailed chemical modelling is required to test this hypothesis.

[C II] 158 $\mu$m emission has been detected by Malfait et al. (1998)
with ISO-LWS and most recently with the PACS instrument on the
Herschel Space Observatory by Sturm et al.\ (2010, submitted). The ISO
fluxes are up to an order of magnitude larger than the model fluxes by
Jonkheid et al. (2007) but a significant fraction of the emission
observed in the large ISO beam may be caused by general Galactic
background emission. More detailed modeling is needed to pin down the
origin of the absence of the [C I] 809 $\mu$m line and check whether
the observed [C II] flux is consistent with enhanced UV radiation.
Future observations of [C I] $J=$1--0 line at 492.161~GHz ($^3$P$_1$--$^3$P$_0$), which is generally brighter than the [C I] $J=$2--1 line, may provide a stronger constraint on the amount of neutral atomic carbon in the disc. Observations of the [C II] line with Herschel-HIFI can directly
resolve the line profile and determine whether this line indeed comes
from the disc. Overall, the [C I] flux and [C I]/[C II] line ratio
could become interesting diagnostics of the UV radiation field to
which the disc is exposed.

\section{Conclusions}\label{s:conclusions}

We summarise our conclusions as follows:

\begin{itemize}
 \item We present evidence for warm molecular gas associated with the disc around HD~100546, in the regions within 400~AU from the star, successfully separated from more extended material in our CHAMP$^+$ observations;
 \item The gas kinematics are consistent with Keplerian rotation around an 2.5~M$_{\odot}$ star of a disc with a 400~AU radius, viewed at an inclination of 50$\degr$ from face-on;
 \item The $^{12}$CO (6--5)/(3--2) line ratio of 1.1$\pm$0.6 is higher than measured towards discs around T Tauri stars, likely due to a more efficient gas heating of the disc containing PAHs by the stronger UV radiation from the B9 star;
 \item {Our data testify to the significant molecular gas reservoir in the disc, consistent with the total disc masses of more than 10$^{-3}$~M$_{\odot}$.}
We exclude the possibility of a low-density disc and optically thin $^{12}$CO emission.
 \item Line asymmetry seen in $^{12}$CO $J=$6--5 and $J=$3--2 lines is explained by a temperature asymmetry, with one side of the disc slightly colder than the other, possibly due to a partial obscuration of one side by a warped inner disc or a high disc rim. We exclude radial asymmetry, midplane density asymmetry and mispointing as possible causes;
\item Our modelling shows that, due to the efficient heating of the disc gas by the star, both low-$J$ and high-$J$ $^{12}$CO lines are dominated by the outermost regions of the disc, though slightly different vertical disc layers with $\Delta T$=15-20~K;
\item {We find that in `colder' discs where temperatures of the emitting regions are close to 20-30~K in the outer disc -- colder Herbig Ae discs and especially T Tauri discs -- the high-$J$ lines probe a larger extent of the disc, starting from as little as 50~AU}; 
 \item The non detection of [C I] $J=$2--1 line may indicate the presence of more carbon-ionising photons than assumed in the B9 model atmosphere. 
\end{itemize}

Future observations with ALMA will be crucial to characterise the disc around HD~100546, and spatially resolve its kinematics and structure. In particular, these observations will allow a detailed comparison between the spatial distribution of the gas traced by the rotational transitions of $^{12}$CO and its isotopologues, and the dust traced with the millimetre continuum emission. Herschel far-infrared data can probe even higher $J$ $^{12}$CO transitions, as well as [O I] and [C II] lines originating from the disc surface. Being a bright, isolated source suspected to harbour a planet, the disc around HD~100546 is one of the prime targets to probe disc structure in the early planet-building phases.

\begin{acknowledgements}
The research of O.~P. and M.~R.~H. is supported through a VIDI grant from the Netherlands Organisation for Scientific Research (NWO). We thank L.~Kristensen for assistance with the reduction of the APEX data, and C.~M.~Wright for useful discussions. We thank the APEX staff for assistance during the observations. Construction of CHAMP$^+$ is a collaboration between the Max-Planck-Institut f\"ur Radioastronomie Bonn, the Netherlands Institute for Space Research (SRON), the Netherlands Research School for Astronomy (NOVA), and the Kavli Institute of Nanoscience at Delft University of Technology, with support from NWO grant 60006331010.
\end{acknowledgements}

\bibliographystyle{aa}
\bibliography{13709refs}

\end{document}